\documentclass[preprint,amsmath,amssymb,nofootinbib]{revtex4}
\usepackage{amsmath}
\usepackage{amssymb}
\usepackage{color}

\newtheorem{definition}{Definition}

\newcommand{\al}{\alpha}
\newcommand{\bal}{{\bar\alpha}}
\newcommand{\bbe}{{\bar\beta}}
\newcommand{\bde}{{\bar\delta}}
\newcommand{\be}{\beta}
\newcommand{\bga}{{\bar\gamma}}
\newcommand{\bka}{{\bar\kappa}}
\newcommand{\blam}{{\bar\lambda}}
\newcommand{\bmu}{{\bar\mu}}
\newcommand{\bnu}{{\bar\nu}}
\newcommand{\bome}{{\bar\omega}}
\newcommand{\bpi}{{\bar\pi}}
\newcommand{\bPsi}{{\bar\Psi}}
\newcommand{\brho}{{\bar\rho}}
\newcommand{\btau}{{\bar\tau}}
\newcommand{\bxi}{{\bar\xi}}
\newcommand{\bX}{{\bar X}}
\newcommand{\bze}{{\bar\zeta}}
\newcommand{\cH}{{\cal H}}
\newcommand{\cL}{{\cal L}}
\newcommand{\de}{\delta}
\newcommand{\ppbz}{{\frac{\partial}{\partial{\bar\zeta}}}}
\newcommand{\ppr}{{\frac{\partial}{\partial r}}}
\newcommand{\ppz}{{\frac{\partial}{\partial\zeta}}}
\newcommand{\bsig}{{\bar\sigma}}
\newcommand{\vps}{{\varepsilon}}
\newcommand{\bvps}{{\bar\varepsilon}}
\newcommand{\heq}{{\hat =}}

\definecolor{lightgray}{rgb}{.7,.7,.7}

\definecolor{red}{rgb}{1,0,0}

\begin{document}

\title{Extremal Isolated Horizon/CFT Correspondence}

\author{Xiao-Ning Wu}\email{wuxn@amss.ac.cn}
\affiliation{Institute of Mathematics, Academy of Mathematics and
System Science, The Chinese Academy of Sciences, Beijing 100190,
China}
\affiliation{Hua Loo-Keng Key Laboratory of Mathematics,
Chinese Academy of Sciences, Beijing 100190, China}

\author{Yu Tian}\email{ytian@gucas.ac.cn}\thanks{Tel:
(86)10-82792321}
\affiliation{College of Physical Sciences, Graduate University of Chinese Academy of Sciences\\
Beijing 100049, China}
\affiliation{Kavli Institute for Theoretical Physics China, CAS\\
Beijing 100190, China}

\date{\today}

\begin{abstract}
The near-horizon limit of the extremal (weakly) isolated horizon is
obtained under the Bondi-like coordinates. For the vacuum case,
explicit coordinate transformation relating the near-horizon metric
under the Bondi-like coordinates and the standard Poincar\'e-type or
global near-horizon metric of the extremal Kerr black hole is found,
which shows that the two geometries are the same. Combined with the
known thermodynamics of the (weakly) isolated horizon, it is argued
that the Kerr/CFT correspondence can be generalized to the case of a
large class of non-stationary extremal black holes.
\end{abstract}

\maketitle

\section{Introduction}

The microscopic origin of the Bekenstein-Hawking entropy of various
black holes is always an interesting problem in theoretical physics.
As should eventually be given by microstate counting in quantum
statistics, this entropy provides important information from the
quantum theory of gravity. In fact, since the first successful
microstate counting of the five-dimensional extremal black holes by
Strominger and Vafa \cite{SV} and the four-dimensional extremal
black holes by Maldacena and Strominger \cite{MS} in the context of
superstring theory, it has been shown by Strominger
\cite{Strominger} that for the BTZ black hole the correct microstate
counting can be done in a dual two-dimensional CFT without appealing
to supersymmetry or string theory, based on the work by Brown and
Henneaux \cite{Brown}.

A recent, essential progress on this subject is the microstate
counting of the four-dimensional extremal Kerr black hole via a dual
chiral two-dimensional CFT living on the boundary of the
near-horizon extremal Kerr (NHEK) geometry \cite{Kerr-CFT}. This
work has been extended to black holes in diverse dimensions and/or
with $U(1)$ gauge symmetries in the framework of
Einstein-Maxwell(-scalar), supergravity or superstring theory
\cite{HHKNT,LMP,AOT,HMNS,CCLP,Nakayama,ITW,PW,CW,Ghezelbash,LMPV,CMN,AS}.
Some related developments can be found in \cite{KK,GG}.
{Furthermore, classification of near-horizon geometries of
stationary extremal black holes, including those with
supersymmetries and/or in higher dimensions also has been carefully
studied, with some powerful theorems proved \cite{KLR,KL}.}

All the known examples, however, of the so-called Kerr/CFT
correspondence are for stationary Kerr-type black holes, i.e. each
of them study black holes in stationary and axial symmetric
spacetime. An interesting question is whether such correspondence is
right for more general black holes, especially for non-stationary
black holes. In order to consider this problem, we need a general
definition of the horizon. Weakly isolated horizon (WIH) is a nice
choice \cite{WIH}. A nice review on the theory of WIH can be found
in \cite{As04}. WIH is a generalization of the Killing horizon of
black hole. It not only covers almost all known Killing horizons,
including Schwarzschild and Kerr horizons, but also contains many
non-stationary cases \cite{Lewandowski}. On the other hand, the
thermodynamics of WIH has been studied extensively, with almost all
elegant results as in the stationary case obtained \cite{thermo}. It
has also been shown that Hawking radiation exists near WIH
\cite{WGH07}. It is then natural and interesting to provide a
microscopic interpretation for the entropy of WIH, possibly by
seeking the generalization of the Kerr/CFT correspondence to the
(weakly) isolated horizon case, which is the main purpose of our
paper. In fact, we will show that this is really possible for the
vacuum, extremal isolated horizon.

This paper is organized as follows. In Sec. \ref{EIH}, we introduce
the general definition of the (weakly) isolated horizon and work out
its near-horizon geometry in the extremal case. In Sec. \ref{CFT},
we take its near-horizon limit. For the vacuum, extremal isolated
horizon, we obtain the explicit form of the near-horizon limit and
find the explicit coordinate transformation to the standard
Poincar\'e-type or global near-horizon metric of the extremal Kerr
black hole, which combined with the known thermodynamics of this
kind of isolated horizon validates the microscopic interpretation
for the corresponding entropy via a dual chiral two-dimensional CFT.
Finally, we conclude and make some discussions in Sec.
\ref{conclude}.

\section{Extremal (Weakly) Isolated Horizon and Its Near-Horizon Geometry}
\label{EIH}

Based on works by A.~Ashtekar and his colleagues \cite{As04}, the
weakly isolated horizon and isolated horizon are defined as
\begin{definition} (weakly isolated horizon)\\
Let $(M, g)$ be a space-time. $\cH$ is a 3-dim null hyper-surface in
$M$ and $l^a$ is the tangent vector filed of the generator of $\cH$.
$\cH$ is said to be a {\bf weakly isolated horizon} (WIH), if
\begin{enumerate}
  \item $\cH$ has the topology of $S^2\times\Re$,
  \item The expansion of the null generator of $\cH$ is zero, i.e.
$\Theta_l=0$ on $\cH$,
  \item $T_{ab}v^b$ is future causal for any future causal vector $v^a$
and the Einstein equation holds in a neighborhood of $\cH$,
  \item $[\cL_l,D_a]l^a=0$ on $\cH$, where $D_a$ is the induced
covariant derivative on $\cH$.
\end{enumerate}
\end{definition}
\begin{definition} (isolated horizon)\\
Let $(M, g)$ be a space-time. $\cH$ is a 3-dim null hyper-surface in
$M$ and $l^a$ is the tangent vector filed of the generator of $\cH$.
$\cH$ is said to be an {\bf isolated horizon} (IH), if
\begin{enumerate}
  \item $\cH$ has the topology of $S^2\times\Re$,
  \item The expansion of the null generator of $\cH$ is zero, i.e.
$\Theta_l=0$ on $\cH$,
  \item $T_{ab}v^b$ is future causal for any future causal vector $v^a$
and the Einstein equation holds in a neighborhood of $\cH$,
  \item $[\cL_l,D_a]=0$ on $\cH$, where $D_a$ is the induced
covariant derivative on $\cH$.
\end{enumerate}
\end{definition}
So the fourth condition for the latter is stronger.

We need to choose coordinates and tetrad to describe their
near-horizon geometry. For the WIH, a convenient choice is the
Bondi-like coordinates. The coordinates are constructed as follows:
The tangent vector of null generator of $\cH$ is $l^a$; Another real
null vector field is $n^a$; The foliation of $\cH$ gives us the
natural coordinates $(\theta,\phi)$; Using a parameter $u$ of $l^a$
and Lie dragging $(\theta,\phi)$ along each generator of $\cH$, we
have coordinates $(u,\theta,\phi)$ on $\cH$; Choosing the affine
parameter $r$ of $n^a$ as the forth coordinate, we obtain the
Bondi-like coordinates $(u,r,\theta,\phi)$ near the
horizon\footnote{The Bondi-like coordinates here are exactly the
so-called Gaussian null coordinates in \cite{KLR,KL}.}. In these
coordinates, we choose the null tetrad as
\begin{eqnarray}
l^a&=&\frac{\partial}{\partial u}+U\frac{\partial}{\partial
r}+X\frac{\partial}{\partial
\zeta}+\bX\frac{\partial}{\partial\bze},\nonumber\\
n^a&=&\frac{\partial}{\partial r},\nonumber\\
m^a&=&\omega\ppr+\xi^3\ppz+\xi^4\ppbz,\nonumber\\
\bar m^a&=&\bome\ppr+\bxi^3\ppbz+\bxi^4\ppz,\label{tetrad}
\end{eqnarray}
where $U=X=\omega=0$ on $\cH$ (following the notation in
\cite{As04}, equalities restricted to $\cH$ will be denoted by
``$\heq$" hereafter), and $(\zeta,\bze)$ are complex coordinates on
the section of $\cH$ (not necessarily $e^{\pm
i\phi}\cot\frac{\theta}{2}$). In words of the Newman-Penrose
formulism, we also require the above tetrad satisfy the following
gauge:
\begin{eqnarray}
&&\nu=\gamma=\tau=\al+\bbe-\pi=\mu-\bmu=0,\nonumber\\
&&\vps-\bvps\ \heq\ \kappa\ \heq\ 0, \label{gaugec}
\end{eqnarray}
which means these tetrad vectors are parallelly transported along
$n^a$ in space-time. Furthermore, the forth requirement in the
definition of WIH implies there exists a one form $\omega_a$ on
$\cH$ such that $D_al^b\ \heq\ \omega_al^b$ and $\cL_l\omega_a\
\heq\ 0$. In terms of the Newman-Penrose formulism, $\omega_a$ can
be expressed as $\omega_a=-(\vps+\bvps) n_a+(\al+\bbe)\bar
m_a+(\bal+\be) m_a$. The above equation means $(\vps+\bvps)$ is
constant on $\cH$. Based on the work of Astekar et.
al.\cite{WIH,As04}, we know the surface gravity of WIH is
$\kappa_t=\omega_al^a$ \cite{As04}, so the extremal condition tells
us that $\vps+\bvps\ \heq\ 0$. In general, the inverse metric has
the following form:
\begin{eqnarray}
\left(g^{\mu\nu}\right)=\left(\begin{array}{cccc}
0&-1&0&0\\
-1&2(|\omega|^2-U)&\bome\xi^3+\omega\bxi^4-X&\bome\xi^4+\omega\bxi^3-\bX\\
0&\bome\xi^3+\omega\bxi^4-X&2\xi^3\bxi^4&|\xi^3|^2+|\xi^4|^2\\
0&\bome\xi^4+\omega\bxi^3-\bX&|\xi^3|^2+|\xi^4|^2&2\bxi^3\xi^4
\end{array}\right).\label{metric}
\end{eqnarray}
It is easy to see that $\sqrt{-g}=\sqrt{h}$ if we denote
$(h_{AB}):=(m^A\bar m^B+\bar m^A m^B)^{-1}$ with $A,B=\theta,\phi$.
Obviously, $h_{AB}$ is the induced metric on the section of WIH.

In order to control the geometry in the neighborhood of the horizon,
we consider the commutative relations of tetrad (the first Cartan
structure equation) \cite{Kr80}. The commutative relation of $l^a$
and $n^a$ is
\begin{eqnarray}
[n,l]=(\gamma+\bga)l+(\vps+\bvps)n-(\btau+\pi)m-(\tau+\bpi)\bar m.
\end{eqnarray}
Using the Bondi gauge (\ref{gaugec}), we get
\begin{eqnarray}
[n,l]=(\vps+\bvps) n-\pi m-\bpi\bar m.\label{cnl}
\end{eqnarray}
The $\partial_r$ component of above equation is
\begin{eqnarray}
\frac{\partial U}{\partial
r}=(\vps+\bvps)-\bpi\bome-\pi\omega,\label{pu}
\end{eqnarray}
So we know $\frac{\partial U}{\partial r}\ \heq\ (\vps+\bvps)$. The
extremal condition tells us that $\partial_rU\ \heq\ 0$. The
$\partial_{\zeta}$ component of (\ref{cnl}) is
\begin{eqnarray}
\partial_rX=-\pi\xi^3-\bpi\bxi^4.\label{pX}
\end{eqnarray}
For non-rotational case, $\pi\ \heq\ 0$, so $\partial_rX\ \heq\ 0$.
In order to get $\partial_r\omega|_{\cH}$, we need to consider the
commutative relation of $n^a$ and $m^a$:
\begin{eqnarray}
[n,m]=\bnu l-(\tau-\bal-\be)n-(\mu-\gamma+\bga)m-\blam\bar m.
\end{eqnarray}
Under the Bondi gauge, its $\partial_r$, $\partial_{\zeta}$ and
$\partial_{\bze}$ components are
\begin{eqnarray}
\partial_r\omega&=&\bpi-\mu\omega-\blam\bome\Rightarrow
\partial_r\omega\ \heq\ \bpi.\label{po}\\
\partial_r\xi^3&=&-\mu\xi^3-\blam\bxi^4,\nonumber\\
\partial_r\xi^4&=&-\mu\xi^4-\blam\bxi^3.\nonumber
\end{eqnarray}
The value of $\xi^3$ and $\xi^4$ are always regular (here regular
means finite and non-zero).

It turns out that, in order to consider the near-horizon limit, we
need the value of $\partial^2_r U$ on horizon. From (\ref{pu}), it
is easy to get
\begin{eqnarray}
\partial^2_r U=(\partial_r\vps+\partial_r\bvps)-(\partial_r\bpi)\bome-\bpi\partial_r\bome-\partial_r\pi\omega-\pi\partial_r\omega.
\end{eqnarray}
On horizon,
\begin{eqnarray}
\partial^2_rU&\heq&(\partial_r\vps+\partial_r\bvps)-\bpi\partial_r\bome-\pi\partial_r\omega\nonumber\\
&\heq&(\partial_r\vps+\partial_r\bvps)-2|\pi|^2.
\end{eqnarray}
So we need the value of $\partial_r\vps$ on horizon. From N-P
equations (the second Cartan structure equation) \cite{Kr80},
\begin{eqnarray}
D\gamma-D'\vps=(\tau+\bpi)\al+(\btau+\pi)\be-(\vps+\bvps)\gamma-(\gamma+\bga)\vps+\tau\pi-\nu\kappa+\Psi_2-\frac{R}{24}+\Phi_{11}.
\end{eqnarray}
Under the Bondi gauge, it becomes
\begin{eqnarray}
-\partial_r\vps=\bpi\al+\pi\be+\Psi_2-\frac{R}{24}+\Phi_{11}.
\end{eqnarray}
so
\begin{eqnarray}
-\partial_r\vps-\partial_r\bvps&=&(\bpi\al+\pi\be+\Psi_2-\frac{R}{24}+\Phi_{11})+(\pi\bal+\bpi\bbe+\bPsi_2-\frac{R}{24}+\Phi_{11})\nonumber\\
&=&2|\pi|^2+2\mathrm{Re}\Psi_2-\frac{R}{12}+2\Phi_{11}.
\end{eqnarray}

Now we consider the $u$-dependence of the near-horizon metric. On
WIH, N-P equations tell us
\begin{eqnarray}
D\al-\bde\vps&=&(\rho+\bvps-\vps)\al+\be\bsig-\bbe\vps-\kappa\lambda-\bka\gamma+(\vps+\rho)\pi+\Phi_{10},\\
D\be-\de\vps&=&(\al+\pi)\sigma+(\brho-\bvps)\be-(\mu+\gamma)\kappa-(\bal-\bpi)\vps+\Psi_1.
\end{eqnarray}
The energy condition of WIH tells us $\Phi_{10}\ \heq\ 0$, and the
Raychaudhuri equation implies
\begin{eqnarray}
\rho\ \heq\ 0,\qquad\sigma\ \heq\ 0,\qquad\Psi_0\ \heq\
0,\qquad\Psi_1\ \heq\ 0.
\end{eqnarray}
The Bondi gauge and WIH condition then implies
\begin{eqnarray}
\partial_u\pi\ \heq\ 0,\qquad\partial_u(\Psi_2-\frac{R}{24})\ \heq\ 0.
\end{eqnarray}
WIH condition also gives
\begin{eqnarray}
\partial_u\xi^3\ \heq\ 0,\qquad\partial_u\xi^4\ \heq\ 0.
\end{eqnarray}
We further impose the stability condition $\mathcal{L}_l\Phi_{11}\
\heq\ 0$, which is automatically satisfied in the vacuum case.

To sum up, we get for WIH
\begin{eqnarray}
U&=&-(2|\pi|^2+\mathrm{Re}(\Psi_2)-\frac{R}{24}+\Phi_{11})r^2+O(r^3),\nonumber\\
\omega&=&\bpi r+O(r^2),\nonumber\\
X&=&-(\pi\xi^3+\bpi\bxi^4)r+O(r^2).
\end{eqnarray}
For the vacuum case, it becomes
\begin{eqnarray}
U&=&-(2|\pi|^2+\mathrm{Re}(\Psi_2)) r^2+O(r^3),\nonumber\\
\omega&=&\bpi r+O(r^2),\nonumber\\
X&=&-(\pi\xi^3+\bpi\bxi^4) r+O(r^2).
\end{eqnarray}
The inverse metric has the following form
\begin{eqnarray}
\left(g^{\mu\nu}\right)&=&\left(\begin{array}{cccc}
0&-1&0&0\\
-1&f_1 r^2+O(r^3)&f_2r+O(r^2)&{\bar f_2}r+O(r^2)\\
0&f_2 r+O(r^2)&2\xi^3\bxi^4&|\xi^3|^2+|\xi^4|^2\\
0&{\bar f_2}r+O(r^2)&|\xi^3|^2+|\xi^4|^2&2\bxi^3\xi^4
\end{array}\right)\label{invers},
\end{eqnarray}
where
\begin{equation}\label{f_1}
f_1=6|\pi|^2+2\mathrm{Re}(\Psi_2)-\frac{R}{12}+2\Phi_{11},\qquad
f_2=2(\pi\xi^3+\bpi\bxi^4)
\end{equation}
are functions of $\theta$ only. So the metric should be
\begin{eqnarray}
\left(g_{\mu\nu}\right)=\left(\begin{array}{ccc}(h_{AB}f_2^A
f_2^B-f_1) r^2+O(r^3)&-1&h_{AB} f_2^B r+O(r^2)\\-1&0&\vec{0}\\h_{AB}
f_2^A r+O(r^2)&\vec{0}&h_{AB}
\end{array}\right),\qquad A,B=\theta,\phi.
\end{eqnarray}
In the above calculation, because $l^a$ and $n^a$ are all chosen as
future pointing, the outside black hole region corresponds to the
region $r<0$. This is inconvenient for later use, so we introduce a
simple coordinate reflection $r\to -r$ and re-express the above
metric in the new coordinates as
\begin{eqnarray}
\left(g_{\mu\nu}\right)=\left(\begin{array}{ccc}(h_{AB}f_2^A
f_2^B-f_1) r^2+O(r^3)&-1&-h_{AB} f_2^B
r+O(r^2)\\-1&0&\vec{0}\\-h_{AB} f_2^A r+O(r^2)&\vec{0}&h_{AB}
\end{array}\right),\qquad A,B=\theta,\phi.\label{m2}
\end{eqnarray}

\section{The Near-Horizon Limit of Extremal Isolated Horizon and Its CFT dual}
\label{CFT}

Under the coordinate re-scaling
\begin{equation}
u=\frac{\tilde u}{\lambda},\qquad r=\lambda\tilde{r}
\end{equation}
and taking the near-horizon limit $\lambda\to 0$
\cite{BH,KLR,Kerr-CFT}, the above metric becomes (after omitting all
the tildes in the expression)
\begin{equation}
ds^2=-f_1 r^2 du^2-2du dr+\hat{h}_{AB} (d\zeta^A-f_2^A r
du)(d\zeta^B-f_2^B r du)
\end{equation}
with $\hat{h}_{AB}=h_{AB}|_\cH$ now functions of $\theta$ only.
{This kind of near-horizon metrics have been obtained in
\cite{KLR,KL} and other papers, but here we achieve this directly
from the viewpoint of extremal isolated horizon, without assuming
any symmetry of the whole space-time.}

The functions $f_1$, $f_2$ and $\hat h_{AB}$ satisfy certain
geometric and dynamic constraints (while having some further gauge
freedom), namely
\begin{eqnarray}
R_{m{\bar m}}=0,\qquad R_{mm}=0.
\end{eqnarray}
These equations can be re-expressed in terms of the inner data
$(\omega_a,\hat h_{ab})$ of the horizon as
\begin{eqnarray}
\widehat{\rm div}{\hat\omega}+\hat\omega^2-K&=&0,\nonumber\\
\nabla_{\bar m}\pi+2\hat\Gamma\pi+\pi^2&=&0,\label{constrain}
\end{eqnarray}
where ``$\hat{\phantom =}$" means projecting on the section $S^2$ of
the horizon, $K$ is the Gaussian curvature of $S^2$, and
${\hat\Gamma}$ the component of connection on $S^2$. These equations
have not yet been solved for the WIH and generic IH \cite{As04}.
However, for the axi-symmetric, electrovac case of extremal IH, J.
Lewandowski and T. Pawlowski have obtained the unique solution for
these functions under certain stability condition for the
electromagnetic field \cite{Le03}, i.e. the functions $f_1$, $f_2$
and $\hat h_{AB}$ (which are intrinsic data on horizon) are unique
for the axi-symmetric, electrovac case.

Concerning the vacuum case, in which the stability condition for the
electromagnetic field is automatically satisfied, the solution of
(\ref{constrain}) is \cite{Le03}
\begin{eqnarray}
\hat{m}^a&=&\frac{1}{2}(P^{-1}\partial_x+i P\partial_\phi),\qquad
x=\cos\theta,\\
\hat{\omega}_a&=&\hat{\star}d{V}+d\ln B
\end{eqnarray}
with (taking the horizon area $A=8\pi$ throughout our calculation)
\begin{equation}
P^2=\frac{1+\cos^2\theta}{2\sin^2\theta},\qquad
{V}=-\arctan\cos\theta,\qquad B=(1+\cos^2\theta)^{1/2}
\end{equation}
and ``$\hat{\star}$" the Hodge star defined by the 2-metric
$\hat{h}$. In fact, we have
\begin{equation}
\hat{h}^{ab}=\hat{m}^a\hat{\bar m}^b+\hat{\bar
m}^a\hat{m}^b=\frac{1}{2}(P^{-2}\partial_x^2+i P^2\partial_\phi^2),
\end{equation}
so
\begin{equation}\label{2-metric}
\hat{h}_{ab}=2(P^2
dx^2+P^{-2}d\phi^2)=2\left(\frac{1+\cos^2\theta}{2}
d\theta^2+\frac{2\sin^2\theta}{1+\cos^2\theta} d\phi^2\right).
\end{equation}
Thus we obtain
\begin{equation}
\hat{\omega}_a=-\frac{2\sin^2\theta}{(1+\cos^2\theta)^2}
d\phi-\frac{\cos\theta\sin\theta}{1+\cos^2\theta} d\theta
\end{equation}
and
\begin{equation}
\pi=-i\nabla_{\bar m} V+\nabla_{\bar m}\ln
B=i\frac{1}{2P}\frac{1}{1+\cos^2\theta}+\frac{1}{2P}\frac{\cos\theta}{1+\cos^2\theta}.
\end{equation}
For vacuum IH we have $\mathrm{Re}(\Psi_2)=-K/2$ with $K$ the
Gaussian curvature of the 2-metric $\hat h$, which can be computed
from (\ref{2-metric}) as
\begin{equation}
K=\frac{2-6\cos^2\theta}{(1+\cos^2\theta)^3}.
\end{equation}
Noting in the vacuum case $R=0=\Phi_{11}$ in (\ref{f_1}), we have
\begin{eqnarray}
  f_1 &=& 6|\pi|^2+2\mathrm{Re}\Psi_2=\frac{1+6\cos^2\theta-3\cos^4\theta}{(1+\cos^2\theta)^3}, \\
  \hat{h}_{AB} f_2^A d\zeta^B &=& 2\hat{\omega}_a=-\frac{4\sin^2\theta}{(1+\cos^2\theta)^2}
d\phi-\frac{2\cos\theta\sin\theta}{1+\cos^2\theta} d\theta,
\end{eqnarray}
so, eventually, the near-horizon limit of vacuum, extremal IH is
\begin{eqnarray}
ds^2&=&-\frac{1+6\cos^2\theta-3\cos^4\theta}{(1+\cos^{2}\theta)^{3}}
r^{2} du^{2}-2du
dr+(1+\cos^{2}\theta)(d\theta+\frac{2r\cos\theta\sin\theta}{(1+\cos^{2}\theta)^{2}}
du)^{2}\nonumber\\
&&+\frac{4\sin^{2}\theta}{1+\cos^{2}\theta}(d\phi+\frac{r
du}{1+\cos^{2}\theta})^{2}.
\end{eqnarray}
This can be shown to be the near-horizon limit of the extremal Kerr
solution. In fact, under coordinate transformation
\begin{eqnarray}
  u &=& t+\frac{1}{\rho}, \\
  r &=& \frac{1+\cos^2\theta}{2}\rho, \\
  \phi &=& \varphi+\frac{1}{2}\ln\rho,
\end{eqnarray}
the above metric becomes
\begin{equation}
ds^{2}=(1+\cos^{2}\theta)(-\frac{\rho^{2}}{4}
dt^{2}+\frac{d\rho^2}{\rho^2}+d\theta^{2})+\frac{4\sin^{2}\theta}{1+\cos^{2}\theta}
(d\varphi+\frac{\rho}{2} dt)^{2},
\end{equation}
which coincides with the near-horizon limit of the extremal Kerr
solution in \cite{BH} with the ADM mass $M=1$ consistent with
$A=8\pi$ in our calculation. Restoring arbitrary $M$ and taking some
trivial coordinate transformations, we obtain exactly the same
Poincar\'e-type or global metrics of NHEK geometry as those in
\cite{Kerr-CFT}.

Now, according to \cite{Kerr-CFT}, this NHEK geometry has an
asymptotic symmetry group (under certain boundary conditions)
containing diffeomorphisms generated by
\begin{equation}\label{diff}
\xi_\epsilon=\epsilon(\varphi)\partial_\varphi-\rho\epsilon'(\varphi)\partial_\rho,
\end{equation}
which form a Virasoro algebra without central charge under Lie
brackets. The conserved charges associated with these diffeomorphism
generators \cite{charge} then form a Virasoro algebra, under Dirac
brackets, with a central charge
\begin{equation}
c_L=\frac{12J}{\hbar}
\end{equation}
with $J=M^2/G$ the angular momentum of the extremal Kerr solution.

On the other hand, thermodynamics of general WIH has been
extensively studied in the literature \cite{thermo}. Ashtekar and
Krishnan give a canonical way to choose the time direction
$\partial_{\hat{t}}$ for general WIH, in order to get meaningful
thermodynamics of the WIH, which actually defines the canonical
coordinates $\hat t$ and $\hat\phi$ coinciding with the
Boyer-Lindquist coordinates in the Kerr case \cite{As04}. Based on
their results, for axial symmetric WIH, the surface gravity
$\kappa_{\hat t}$, horizon angular velocity $\Omega_{\hat t}$ and
horizon mass $M_{\hat t}$ associated with the canonical time also
coincide with the corresponding quantities in the Kerr case. It has
been shown that Hawking radiation exists near general WIH
\cite{WGH07}. The temperature of this thermal radiation is
$T=\frac{\kappa_{\hat t}}{2\pi}$, so the Boltzmann factor observed
by a canonical observer should be
\begin{eqnarray}
{\exp(-\frac{E-m\Omega_{\hat t}}{T})}\ ,
\end{eqnarray}
where $E$ is the particle energy and $m$ the particle angular
momentum associated with the canonical observer. We also introduce
coordinates $t,\phi$ as
\begin{eqnarray}
{t}=\frac{\lambda\hat t}{2M_{\hat t}},\qquad
{\phi}=\hat{\phi}-\Omega_{\hat t}\hat{t},
\end{eqnarray}
which coincide, in the extremal case, with the coordinates $t,\phi$
in \cite{Kerr-CFT}. Then the particle energy and angular momentum
associated with the new observer are
\begin{eqnarray}
n_R=(2M_{\hat t} E-m),\qquad n_L=m,
\end{eqnarray}
so we can reexpress the Boltzmann factor as
\begin{eqnarray}
\exp(-\frac{E-m\Omega_{\hat
t}}{T})=\exp(-\frac{n_L}{T_L}-\frac{n_R}{T_R}).
\end{eqnarray}
Under the extreme limit $J\to M^2/G$, it is easy to get
\begin{eqnarray}
T_L=\frac{1}{2\pi},\quad T_R=0.
\end{eqnarray}
This means that the left modes are then thermally distributed with
the temperature $\frac{1}{2\pi}$.

Upon applying the Cardy formula, the microscopic entropy for the CFT
dual to the vacuum, extremal IH is
\begin{equation}
S_{\rm CFT}=\frac{\pi^2}{3} c_L T_L=\frac{2\pi J}{\hbar},
\end{equation}
which is the same as the macroscopic entropy $S_{\rm IH}$ of the
vacuum, extremal IH \cite{As04}.

\section{Concluding Remarks}
\label{conclude}

In this paper, we show that the entropy of the vacuum, extremal
isolated horizon may have microscopic origin, by working out its
near-horizon limit, finding the explicit coordinate transformation
to the standard Poincar\'e-type or global near-horizon metric of the
extremal Kerr black hole, and using the original Kerr/CFT argument.
This supports the generalization of the Kerr/CFT correspondence to
the extremal isolated horizon/CFT correspondence. Based on our
discussion, the near horizon geometry of IH is fixed by the inner
data of the horizon, i.e. $(h_{ab},\omega_a)$. This may tell us that
the origin of black hole entropy is only related with the inner
geometry of the horizon and does nothing with how we embed it into
the space-time, at least in the framework of Einstein gravity.

The near-horizon limit that we obtain is for the general weakly
isolated horizon, but only for the special case of the vacuum,
extremal isolated horizon exact solution of the intrinsic data on
the horizon is known. It is expected that more general cases can be
dealt with, when there is no explicit solution and more general
boundary conditions for the asymptotic symmetry of the near-horizon
geometry have to be found. This is left for future work.

\begin{acknowledgments}
We thank Dr. W. Song, Prof. C.-G. Huang and Prof. Y. Yang for
helpful discussions. This work is partly supported by the National
Natural Science Foundation of China (Grant Nos. 10605005, 10705048,
10605006 and 10731080) and the President Fund of GUCAS.
\end{acknowledgments}

\end{document}